\documentclass[seceq]{ptptex}
\usepackage{graphicx} 
\usepackage{latexsym}
\notypesetlogo
\begin{document}
\begin{flushright}
RUP-16-4 \\
\end{flushright}
\vspace{10mm}
\begin{center}
\Large{ \bf Schwinger-Dyson equation in Minkowski space beyond the IE approximation}
\vspace{15mm}

\large{ Shuji {\sc Sasagawa} and Hidekazu {\sc Tanaka}\\
Department of Physics, Rikkyo University, 
           Nishi-ikebukuro, Toshima-ku Tokyo, Japan, 171\\
}
 \end{center}

\begin{center}

\vspace{25mm}

{\Large ABSTRACT}
 \end{center}
        
\vspace{10mm}
\def\proj{{\bf P}}  
\def\slsh#1{{#1}{\kern-6pt}/{\kern1pt}}  



        
We investigate the properties of fermion mass functions in quantum electrodynamics calculated by the Schwinger-Dyson equation in the strong coupling region, in which the loop integration is performed in Minkowski space. The calculated results without the instantaneous exchange approximation are compared with those obtained by integration in Euclidean space. 

\newpage

\section{Introduction}

 Chiral phase transition has been studied by various methods. One such method is implementation of the Schwinger-Dyson equation (SDE) [1,2], which can evaluate nonperturbative phenomena. So far, many works for chiral symmetry breaking have been done with the SDE in momentum representation, in which a one-loop contribution is integrated over Euclidean space.  

Some calculations of fermion mass functions  with the SDE have been done in Minkowski space.  In Ref. [3], spectral representation for Green functions is assumed, in which the mass functions are calculated in Lorentz-invariant form.  In Ref. [4], explicit one-loop contributions of the mass function have been calculated. However the mass function is evaluated only one iteration from a constant initial mass as an input. 

In order to extend the SDE at finite temperature, we need to integrate the energy and the momentum separately, due to existence of the Boltzmann factor.
 In Ref. [5], the mass function is analyzed in Euclidean space at zero temperature, with the energy and momentum integrated separately.

At finite temperature for equilibrium systems, the imaginary-time formalism (ITF) is implemented, which continues to Euclidean space at the zero temperature limit.

On the other hand, the real-time formalism (RTF) for nonequilibrium systems is formulated in Minkowski space. The SDE in RTF has been studied with the instantaneous exchange approximation (IEA) [6,7] in which gauge boson energy is neglected. In the IEA, the mass function does not depend on the energy. Furthermore, the critical coupling of chiral symmetry breaking in IEA is about half of that calculated with four momentum integration in Euclidean space at zero temperature [7-10].  An alternative method has been proposed in Ref. [11], in which an energy independence for the mass function is imposed. In evaluation of the mass function, only results for the static limit as momentum $p\rightarrow 0$ with zero energy $p_0=0$  have been presented [12].
 
 Analytic continuation from Euclidean space to Minkowski space is valid in perturbative calculation if pole positions in the complex plane of energy are known. However it is not trivial in the nonperturbative region.

So far, the structure of the fermion mass function in the strong coupling region in the entire range of energy and momentum space has not been fully studied in Minkowski space, even at zero temperature.

In this paper, we study the fermion mass function with the SDE in Minkowski space beyond the IEA in Abelian gauge theory, such as quantum electrodynamics (QED). Before studying it at finite temperature, we first study the  properties of the fermion mass function in  energy and momentum space at zero temperature.

In Sect. 2, we formulate the SDE in Minkowski space. In Sect. 3, some numerical results are shown and calculated results are compared with those obtained by the SDE in Euclidean space.   Section 4 is devoted to a summary and some comments. 

\section{The SDE for the fermion mass function }

We calculate a fermion self-energy $\Sigma(P)$ in QED in 4-dimensions, which is given by
\begin{eqnarray}
 -i\Sigma(P)=(-ie)^2\int{d^4Q \over (2\pi)^4}\gamma^{\mu}iS(Q)\Gamma^{\nu}iD_{\mu\nu}(K),
\end{eqnarray}
where $S(Q)$ and $D_{\mu\nu}(K)$ are propagators of a fermion with momentum $Q=(q_0,{\bf q})$ and a photon with momentum $K=P-Q=(k_0,{\bf k})$, respectively, 
Here, $P=(p_0,{\bf p})$ is an external momentum of the fermion. 

The fermion propagator is given by
\begin{eqnarray}
iS(Q) = {iZ \over \slsh{Q}-\Sigma(Q)+i\varepsilon}={i \over A(Q)\slsh{Q}-B(Q) +i\varepsilon }.
\end{eqnarray}
 In this paper, we calculate a mass function $M={\rm tr}(\Sigma)/4$  in the Landau gauge, in which the wave-function renormalization constant is $Z=1$  in a one-loop order of perturbation. Therefore, we calculate the self-energy given in Eq.(2$\cdot$1) with $A=1$, $\Sigma=B$, and the fermion-photon  vertex with $\Gamma_{\mu}=\gamma_{\mu}$. Here, the photon propagator is given as 
\begin{eqnarray}
iD_{\mu\nu}(K)=\left(-g_{\mu\nu}+{K_{\mu}K_{\nu} \over K^2}\right){i \over K^2+i\epsilon}
\end{eqnarray}
in the Landau gauge.  Integrating over the azimuthal angle of the momentum ${\bf q}$, the mass function is given by 
\begin{eqnarray}
M_M(p_0,p) = -{3i\alpha \over 2\pi^2}\int^{\Lambda_0}_{-\Lambda_0}dq_0 \int^{\Lambda}_{\delta} dq {q \over p}[M_MIJ](q_0,q) 
\end{eqnarray}
with $p=|{\bf p}|$ and $q=|{\bf q}|$. For $I,M_M$ and $J$,  we separate the real parts $I_R,M_R,J_R $ and the imaginary parts $I_I,M_I,J_I$, respectively. Here,$I$ and $J$ are given by
\begin{eqnarray}
I= {1 \over Q^2-M_M^2+i\varepsilon} ={q_0^2-(E^2)_R+i(E^2)_I \over (q_0^2-(E^2)_R)^2+((E^2)_I)^2},
\end{eqnarray}
where 
\begin{eqnarray}
 (E^2)_R=q^2+(M^2)_R 
\end{eqnarray}
and
\begin{eqnarray}
 (E^2)_I=(M^2)_I-\varepsilon,
\end{eqnarray}
respectively, with 
\begin{eqnarray}
 (M^2)_R=Re(M_M^2)=(M_R)^2-(M_I)^2 
\end{eqnarray}
and
\begin{eqnarray}
 (M^2)_I=Im(M^2_M)=2M_RM_I.
\end{eqnarray}

The photon propagator is given by 
\begin{eqnarray}
J=\int^{\eta_+}_{\eta_-}dk{k \over K^2+i\varepsilon}
\end{eqnarray}
with $\eta_{\pm}=|p\pm q|$ and $k=|{\bf k}|$.

For $J$, we can integrate over $k$ as
\begin{eqnarray}
 J_R = -\int^{\eta_+}_{\eta_-}dk{k(k^2-k_0^2) \over (k^2-k_0^2)^2+\varepsilon^2} =  -{1 \over 4}\log{(\eta_+^2-k_0^2)^2+\varepsilon^2 \over (\eta_-^2-k_0^2)^2+\varepsilon^2} 
\end{eqnarray}
 and
$$ J_I= -\int^{\eta_+}_{\eta_-}dk{k\varepsilon \over  (k^2-k_0^2)^2+\varepsilon^2} = -{1 \over 2}\left[\arctan{\eta_+^2-k_0^2 \over \varepsilon}-\arctan{\eta_-^2-k_0^2 \over \varepsilon}\right] $$
\begin{eqnarray}
\rightarrow -{\pi \over 4}\left[\epsilon\left(\eta_+^2-k_0^2\right)-\epsilon\left(\eta_-^2-k_0^2\right)\right] 
\end{eqnarray}
for $\varepsilon \rightarrow 0$.
 Here, we define $\epsilon(z)=\theta(z)-\theta(-z)$ with the step function $\theta(z)$.

In this paper, we approximate the integration over $q_0$ as
\begin{eqnarray}
M_M(p_0,p) \simeq -{3i\alpha \over 2\pi^2} \int^{\Lambda}_{\delta} dq {q \over p}\sum_{l=1}^{N-1}<[M_MJ](q)>_l I(q_0^{(l+1)},q_0^{(l)})
\end{eqnarray} 
with
\begin{eqnarray}
I(q_0^{(l+1)},q_0^{(l)}) = \int^{q_0^{(l+1)}}_{q_0^{(l)}}{dq_0 \over Q^2-M_M^2+i\varepsilon}.
\end{eqnarray}
Here, the real  and imaginary parts of the fermion propagator are given by
$$ 
I_R^{(l)}=ReI(q_0^{(l+1)},q_0^{(l)}) ={\epsilon(<2q_0-{\partial (E^2)_R\over\partial q_0}>_l) \over 2<|2q_0-{\partial (E^2)_R \over\partial q_0}|>_l } 
$$
\begin{eqnarray}
\times \log{[(q_0^{(l+1)})^2-<(E^2)_R>_l]^2+<(E^2)_I>_l^2 \over [(q_0^{(l)})^2-<(E^2)_R>_l]^2+<(E^2)_I>_l^2} 
\end{eqnarray}
and
$$ I_I^{(l)}=ImI(q_0^{(l+1)},q_0^{(l)})={\epsilon(<2q_0-{\partial (E^2)_R\over\partial q_0}>_l)\epsilon(<(E^2)_I>_l) \over <|2q_0-{\partial (E^2)_R \over\partial q_0}|>_l } $$
\begin{eqnarray}
 \times \left[\arctan{(q_0^{(l+1)})^2-<(E^2)_R>_l \over |<(E^2)_I>_l|}-\arctan{(q_0^{(l)})^2-<(E^2)_R>_l \over |<(E^2)_I>_l|}\right], 
\end{eqnarray}
respectively.  Here, $<X>_l$ denotes an average of  $X(q_0^{(l+1)})$ and $X(q_0^{(l)})$ as $ <X>_l=[X(q_0^{(l+1)})+X(q_0^{(l)})]/2$.

In Euclidean space, the SDE for the mass function is given by
\begin{eqnarray}
 M_E(p_4,p)={3\alpha \over 4\pi^2}  \int^{\Lambda_4}_{-\Lambda_4} dq_4\int^{\Lambda}_{\delta} dq{q \over p}\log\left({k^2_4+\eta^2_+ \over k_4^2+\eta_-^2}\right){{M_E}(q_4,q) \over Q_E^2+\left(M_E(q_4,q)\right)^2}
\end{eqnarray}
with $k_4=p_4-q_4$ and $Q_E^2=q_4^2+q^2$, which has been studied in Ref. [5].

In the IEA, setting $k_4=0$ and integrating over $q_4$, the mass function is given by 
\begin{eqnarray}
M_{IEA}(p)={3 \alpha \over 2\pi}\int^{\Lambda}_{\delta}dq{q \over p}M_{IEA}(q)\log\left({\eta_+^2\over \eta_-^2}\right){1 \over 2E}
\end{eqnarray}
with $E=\sqrt{q^2+(M_{IEA})^2}$, which is the same expression as  the mass function in Minkowski space with $Im M_{IEA} =0$.

\section{Numerical results}

In this section, some numerical results are presented. 

 We solve the SDE  by a recursion method starting from a constant mass.  

The mass function can be written as an absolute value $|M_M(p_0,p)|$ and a phase factor  $\exp(i\Phi(p_0,p))$ as 
\begin{eqnarray}
M_M(p_0,p)=|M_M(p_0,p)|\exp(i\Phi(p_0,p)).
\end{eqnarray}
In order to search for a critical point in which the chiral symmetry is broken, we evaluate 
\begin{eqnarray}
<M_M>=\int^{\Lambda_0}_{-\Lambda_0}dp_0\int^{\Lambda}_{\delta}dp|M_M(p_0,p)|,
\end{eqnarray}
\begin{eqnarray}
<M_E>=\int^{\Lambda_4}_{-\Lambda_4}dp_4\int^{\Lambda}_{\delta}dpM_E(p_4,p)
\end{eqnarray}
and 
\begin{eqnarray}
<M_{IEA}>=2\Lambda_0\int^{\Lambda}_{\delta}dpM_{IEA}(p).
\end{eqnarray}
as order parameters.
We take $\Lambda_0=\Lambda_4=4\Lambda$ and $\delta=0.01\Lambda$ in energy and momentum integrations, respectively.

In Fig. 1, the convergence property of $<M_M>/\Lambda$  at $\alpha=2$ is presented by a solid line. The horizontal axis denotes the number of iterations. The $+$ symbols and the dotted line represent calculated results for $<M_E>/\Lambda$ and $<M_{IEA}>/\Lambda$, respectively. As shown in Fig. 1,  the mass functions for the three cases  rapidly converge.
\footnote{Initial input parameters are $M_R=0.5\Lambda$ and $M_I=0$.}
\begin{figure}
\centerline{\includegraphics[width=10cm]{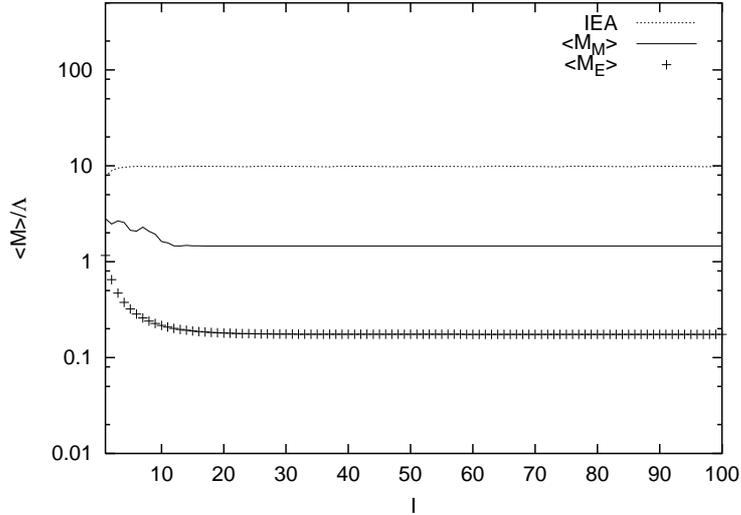}}
\caption{The solid line represents the convergence property of $<M_M>/\Lambda$ at  $\alpha=2$. The horizontal axis denotes the number of iterations. The $+$ symbols and the dotted line denote calculated results for $<M_E>/\Lambda$ and $<M_{IEA}>/\Lambda$, respectively.}
\end{figure}

In Fig. 2, dependences on the coupling constant $\alpha$ for $<M_i>/\Lambda~~(i=M,E,IEA)$  are presented.  The critical coupling for the case of $<M_M>/\Lambda$ is close to that for the case of $<M_E>/\Lambda$, though the two cases have different $\alpha$ dependences above the critical points. On the other hand, the critical coupling for the case of the IEA is much smaller\footnote{The critical point calculated  with the IEA is consistent with the result obtained in Ref. [7] at $T=0$.}than those for the other two cases.

\begin{figure}
\centerline{\includegraphics[width=10cm]{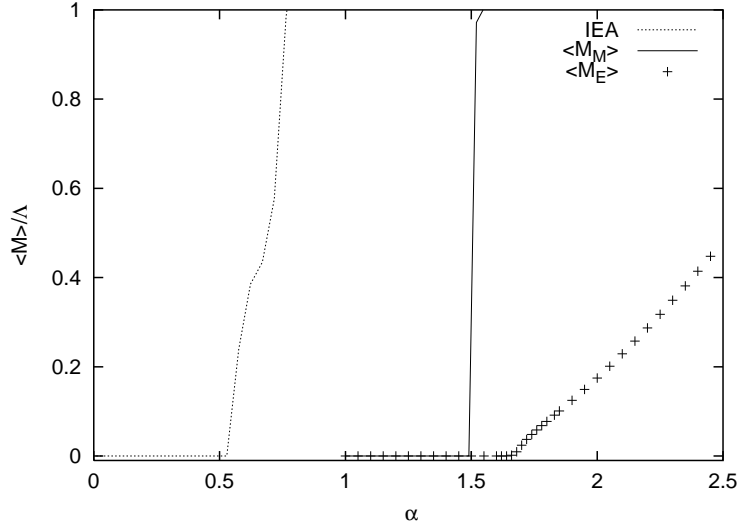}}
\caption{ The solid line represents the $\alpha$ dependence of $<M_M>/\Lambda$ . The $+$ symbols and the dotted line denote the results for $<M_E>/\Lambda$ in Euclidean space and the case of the IEA, respectively.}
\end{figure}

In Fig. 3,  $p_0$ dependences of $|M_i|/\Lambda~~(i=M,E,IEA)$ integrated over the range of $0.01\Lambda \leq p \leq \Lambda$  are  presented at $\alpha=2$.   As shown in the figure, the mass function calculated in Minkowski space without the IEA depends on $p_0$.

 The mass function with a restriction of $|E_I|/|E_R| < 1$ is shown by a dashed line.\footnote{ We have that $E=E_R+iE_I$ is given by $E_R=ReE=\sqrt{|E^2|}\cos(\Phi_E/2)$ and $E_I=ImE=\sqrt{|E^2|}\sin(\Phi_E/2)$, with $|E^2|=\sqrt{((E^2)_R)^2+((E^2)_I)^2}$. The phase $\Phi_E$ is defined as $\Phi_E=\arctan((E^2)_I/(E^2)_R)$. Here, the angle is defined in the region  $-\pi/2 \leq \Phi_E \leq \pi/2$ for $(E^2)_R \geq 0$. For $(E^2)_R < 0$ , $\pi/2 < \Phi_E  \leq \pi$ for $(E^2)_I \geq 0$ and $-\pi < \Phi_E  < -\pi/2$ for $(E^2)_I < 0$. }
The imaginary part of the energy denoted by $E_I$ relates instability, such as decay properties, for the fermion with an effective mass. In the space-like momentum region, there may be no $E_I$ as expected from perturbative calculations. For small $p_0$, for which contributions from space-like momentum are important, the constraint $|E_I|<|E_R|$ is already satisfied. On the other hand, for large $p_0$, for which the contributions from time-like momentum are important, the imaginary part of $E$ is expected to be generated.
In the region rejected  by the constraint $|E_I|<|E_R|$, which corresponds to the region above the dashed line in Fig. 3, the massive fermion states may be unstable due to large $|E_I|$.

  The mass function $|M_E|/\Lambda$ integrated over the range  $0.01\Lambda \leq p \leq \Lambda$ in Euclidean space is shown by $+$ symbols.  
  
The $p_0$ dependence of the mass functions are similar shapes in both cases, though the mass function in Minkowski space is much larger than that in Euclidean space at fixed $\alpha$ due to  different $\alpha$ dependence above the critical coupling.

\begin{figure}
\centerline{\includegraphics[width=10cm]{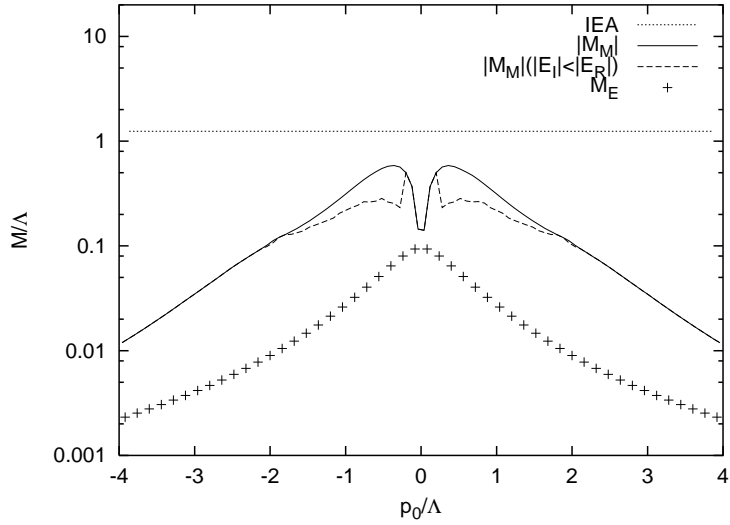}}
\caption{The $p_0$ dependence of $|M_M|/\Lambda$ integrated over $0.01\Lambda \leq p \leq \Lambda$ at $\alpha=2$ is presented by the solid line.  The $+$ symbols and the dotted line represent the calculated  results for $M_E/\Lambda$ and $M_{IEA}/\Lambda$, respectively. The dashed line denotes $|M_M|/\Lambda$ with $|E_I| < |E_R|.$}
\end{figure}

In Fig.4, the mass function $|M_M(p_0,p)|/\Lambda$ with $\alpha=2$ in ($p_0/\Lambda$, $p/\Lambda$) space is shown. The structure of the mass function in Minkowski space is rather more complicated than that in Euclidean space [5].

\begin{figure}
\centerline{\includegraphics[width=15cm]{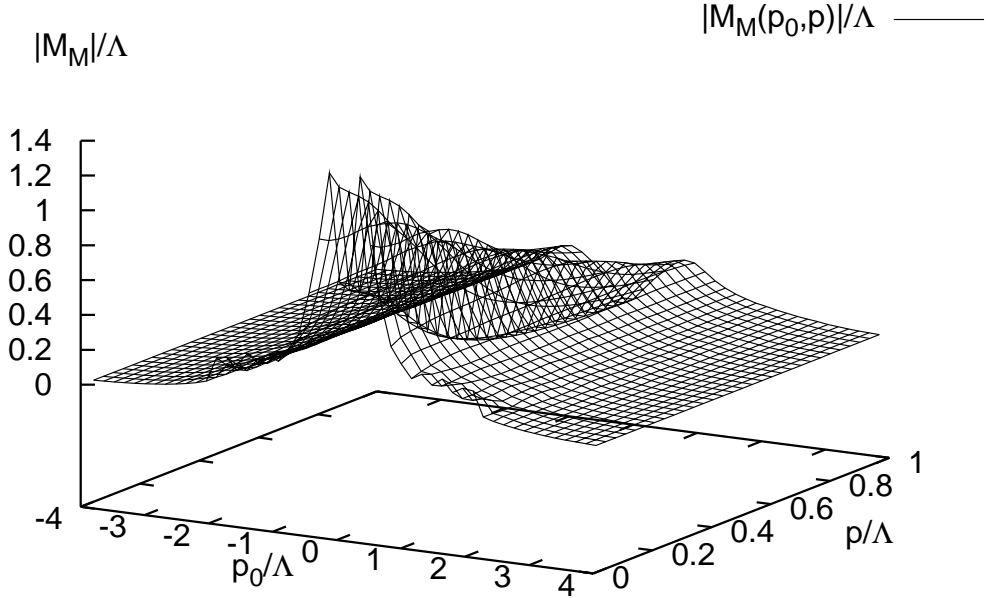}}
\caption{Three-dimensional plot of the mass function $|M_M(p_0,p)|/\Lambda$ with $\alpha=2$. }
\end{figure}

\section{Summary and Comments}

In this paper, we studied a fermion mass function  solved by the Schwinger-Dyson equation (SDE) in Minkowski space  beyond the instantaneous exchange approximation (IEA). 

Beyond the IEA, the mass function depends on a zero component of the fermion momentum $p_0$, as well as that of the space component $p$. 

In order to study phase transition, we defined integrated mass functions as order parameters and examined chiral symmetry breaking in Minkowski space as well as that in Euclidean space.
Here, the mass functions are integrated over the ranges $-4\Lambda \leq p_0 \leq 4\Lambda$ and $\delta \leq p \leq \Lambda$, respectively, with cut-off parameters $\Lambda$ and $\delta$. 

 We found that the critical coupling for the mass function  in Minkowski space is close to that in Euclidean space, though the two cases have different dependences for the coupling constant $\alpha$ above the critical couplings.
On the other hand, the critical coupling for the case of the IEA is much smaller than those for other two cases. 

 Furthermore, the $p_0$ dependences of the mass functions integrated over the momentum $p$ are also similar in shape in both cases, though the mass function in Minkowski space is much larger than that in Euclidean space at fixed $\alpha$ due to the different $\alpha$ dependence above the critical coupling.

Evaluation of the mass functions in Minkowski space allow us to study instability of the fermion with an effective mass, which depends on an imaginary part of the energy for the massive fermion states.  In our calculation, for small $p_0$, for which contributions from space-like momenta are important, the constraint  $|E_I|<|E_R|$ is satisfied. Here, $E_R$ and $E_I$ denote the real  and  imaginary parts of the energy for the massive fermion states, respectively.
  In the region rejected  by the constraint $|E_I|<|E_R|$, the fermion with  effective mass may be unstable due to large $E_I$.

We also presented the structure of the mass function in $(p_0/\Lambda,p/\Lambda)$ space in the strong coupling region, which may be useful information for evaluation of processes that include fermions with an effective mass in the strong coupling region in Minkowski space.  

It may be expected that the SDE has multiple solutions in numerical calculations.  We should identify physical solutions among them, such as  a solution  with the lowest energy. Further studies are needed for the phase factor in order to know the behaviors of the real and imaginary parts of the mass function separately in  Minkowski space in the strong coupling region.

In future works, we shall  extend our method to QCD and  finite temperature with the real-time formalism.

\section*{Acknowledgements}

This work was partially supported by MEXT-Supported Program for the Strategic Research Foundation at Private Universities, 2014-2017.






\end{document}